\documentclass{ifacconf}
\usepackage{natbib}  
\usepackage{graphicx} 

\usepackage{tabularx}

\begin{document}
\begin{frontmatter}
\thanks[footnoteinfo]{This research was supported by the University of Houston through
a start-up grant.}
\title{Neurological Status Classification Using Convolutional Neural Network} 


\author[First]{Mehrad Jaloli} 
\author[Second]{Divya Choudhary} 
\author[First]{Marzia Cescon}

\address[First]{Dept. of
Mechanical Engineering, University of Houston, Houston, TX United States 77004}
\address[Second]{Dept. of Biochemistry, University of Oxford, South Parks Road, Oxford, OX1 3QU, UK}
\vspace{2pt}
(Corresponding e-mail: mcescon2@uh.edu)

\begin{abstract}                

In this study we show that a Convolutional Neural Network (CNN) model is able to accurately
discriminate between 4 different phases of neurological status in a non-Electroencephalogram
(EEG) dataset recorded in an experiment in which subjects are exposed to physical, cognitive
and emotional stress. We demonstrate that the proposed model is able to obtain 99.99\% Area
Under the Curve (AUC) of Receiver Operation characteristic (ROC) and 99.82\% classification accuracy on
the test dataset. Furthermore, for comparison, we show that our models outperforms traditional
classification methods such as SVM, and RF. Finally, we show the advantage of CNN models, in comparison to other methods, in robustness to noise by 97.46\% accuracy on a noisy dataset.

\end{abstract}


\end{frontmatter}

\section{Introduction}

Physiological changes in the human body are accompanied by, or are a results of, changes in neurological status. In conditions such as diabets [\cite{farrell2004impact}] or seizures [\citet{mueller1979effect}], it has been shown that a surge in stress levels was linked to a deterioration in patients health condition. Recent improvements at the interface of healthcare and technology allowed to track neurological changes during arousal/stress and infer the subject's underlying neurological state at each time instant. However, most of those studies have used neurological datasets like electroencephalogram (EEG) [\citet{al2015mental}] or functional magnetic resonance imaging (fMRI) [\citet{dopfel2018mapping}], which 
mostly have invasive, uncomfortable and time-consuming data collection protocols.

With the advent of non-invasive sensors built-in devices and the rapid growth in the adoption of wearable devices, it is now possible to collect  different types of physiological signals that can be leveraged to monitor internal physiological changes. Some of this signals that may be useful in studying the neurological status are electrodermal activity (EDA), heart rate (HR), accelerometer (ACC), skin temperature (Temp) and respiratory rate [\citet{cogan2014wrist}].

In particular, previous studies on EDA and HR have shown that features extracted in both frequency and time domain are relevant for neurological status analysis. Posada-Quintero and coworkers showed in  [\citet{posada2018time}] how the EDA signal dynamics shift in various ranges of frequency and \citet{ghiasi2020assessing} showed how heart rate along with EDA signal contributes in detecting the arousal levels.
Such studies have inspired researchers to build more analytically powerful tools and methods to accurately predict the intensity and type of the stress and also qualitatively analyse its correlation with physiological changes and external stimuli.

Against this background, our goal is to develop and verify a method which is able to  discriminate between different stages of neurological status and arousal in a non-EEG dataset. 
The outline of the paper is as follows. Section 2 introduces some background on deep neural networks; in section 3 the dataset is presented followed by the explanation of the preprocessing steps. Sections 4 and 5 explain the results and conclusions, respectively;

\section{Deep Neural Network } \label{NN}

Artificial neural networks (ANN) are categorized as a type of machine learning algorithms inspired by the human brain, consisting of a stack of neurons, called layers, such that the neurons within the same layer are not connected to each other. Deep neural networks, in particular are types of ANNs with multiple layers connecting input and output layers, called hidden layers [\citet{lecun2015deep}].
The input layer, receives the input data, the hidden layers are where the input is processed and the output layer provides the output of the model.

\subsection{Convolutional Neural Network (CNN)}
CNN is a well-known deep learning architecture that has demonstrated an excellent performance in computer vision and image classification problems [\citet{Karpathy_2014_CVPR}].
It has been proven to have many advantages over other machine learning algorithms like robustness to noise, automatic feature extraction, weight sharing characteristic, being suitable for working with incomplete patterns and flexible structures which makes them a compatible and proper tool for facing with different size of datasets and complicated problems.

A CNN model is generally composed of two parts: 1) feature extraction, which includes several convolution layers followed by pooling layers and activation functions and 2) classification part that is consisted of fully connected layers, structured the same as an ANN, which exploits the extracted features from the first part and performs the classification task. 

Generally, there are three well-known CNN structures, all can be used in classification problems: 1 dimensional CNN (1D CNN), 2 dimensional CNN (2D CNN) and 3 dimensional CNN (3D CNN). 1D CNN is able to derive features from fixed-length slices of a dataset, by sliding 1 dimensional convolution filters all along each slice (window) such that the width of the filter is equal to the width of the dataset. 
Such properties makes 1D CNN a very powerful and fast method in identifying simple patterns within a data which can be used for more complex patterns in higher layers of the model.

2D and 3D CNNs, have the same aforementioned characteristics, while the difference is in the dimensionality of the input data and convolutional kernels. In 2D CNN , 2D filters slide in both width and height dimensions of the data segment. It is mostly applied on image datasets, where the spatial information of the features is of interest in addition to features by its own [\citet{lecun2015lenet}]. However, 3D CNN in which 3D kernels move through 3 dimensions of the data, height, width and depth, is mostly used in video datasets, due to the capability of extracting features from both the spatial and the temporal dimension of the segments of the dataset [\citet{ji20123d}].

Fig \ref{cnn-arch} shows a sample CNN model, where the input is an image of size $H_I \times W_I \times D_I$, with $H_I, W_I$ and $D_I$ being height, width and depth, or number of  color channels, of the input image, respectively. The image is turned to a tensor of raw pixels, to be able to fed and analyzed by a CNN model.  

In a 2D CNN, filters of size $H_f, W_f$ slide all over the tensors in a separate manner and result a deeper tensor called feature map. In cases that we have no padding , the size of the output of each convolution layer with stride 1 is $H_{conv} \times W_{conv} \times D_f$ such that: 

\begin{equation}
    H_{conv}= H_{I} - H_{f} +1 
\end{equation}    
\begin{equation}
 W_{conv}= W_{I} - W_{f} +1 
\end{equation}    

 and $D_f$  is the number of filters of a convolution layer.  $W_{I},~ W_{conv},~ H_{I},~ H_{conv} $ are the width and height of the convolution layer input and output, respectively. $H_{f}$ and $W_{f}$ are height and width of the filter.
 Note that in 1D CNN, the width of the filter, $W_f$, is equal to the width of data segments, $W_I$.
 Afterward, to apply a nonlinear transformation, the output of the convolution layer is sent to an activation function. Some common activation functions are rectified linear unit (RELU), tangent hyperbolic (TanH), exponential Linear units (ELU), softmax. 
 Table \ref{table:activation} lists the corresponding mathematical expressions.
 Note that in ELU equation, $a\geq 0$ is a hyper-parameter that needs to be tuned.

\begin{figure}[b]
 \centering 
 \includegraphics[width=\columnwidth]{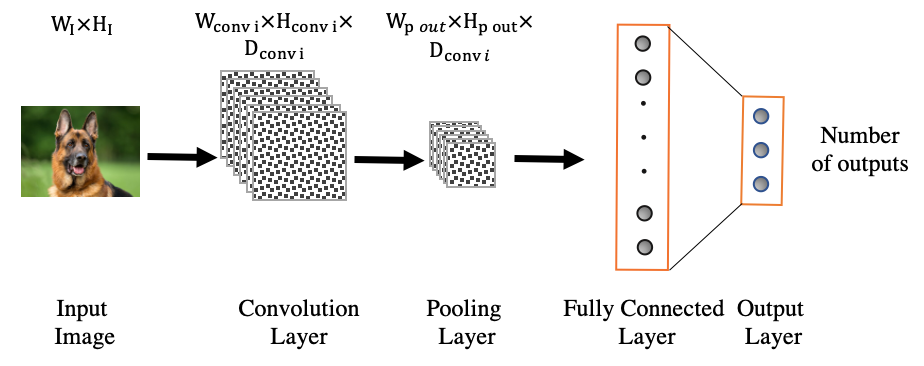}
\caption{Illustration of the CNN architecture. Convolutional layer, pooling layer and fully connected layer are the core CNN operation and output layer is final output.\protect}
 \label{cnn-arch\protect }
\end{figure}

\begin{table}[h]
\centering
\caption{Activation Functions\protect}
\begin{tabular}{cc}

Activation Function & Equation \\
\hline
\vspace{5pt}

RELU & $ f(x)$ =$ \left\{
    \begin{array}{ll}
     x ~ for ~ x>0\\
    0 ~ for ~ x<0
    \end{array}
\right.$ \\

\vspace{5pt}

TanH & $ f(x)$ = $\frac{e^x - e^{-x}}{e^x + e^{-x}}$\\ 
     
\vspace{5pt}

ELU & $ f(a,x) = \left\{
    \begin{array}{ll}
        x ~~~~~~~~~~~~ for   ~ x\geq 0\\
        a(e^x - 1) ~ for ~ x<0
    \end{array}
\right.$ \\

Softmax & $f(x) =$ $\frac{1}{1+ e^{-x}}$ \\
\hline  
\end{tabular}
\label{table:activation}

\end{table}

In order to reduce the amount of parameters and computational load and avoid over-fitting, feature maps are sub-sampled into lower-dimensions using pooling layers.
Two of the most common pooling layers are Maxpooling and Average pooling. 
By applying max/average pooling function of size $H_{p}$ , $W_{p}$, the max/average of each region of size ($H_{p}$, $W_{p}$) is calculated and outputs a subsampled feature map with size:

\begin{equation}
    W_{p ~out}= (W_{conv}- W_{p})/s+1 
\end{equation}
\begin{equation}
    H_{p~ out}=(H_{conv}-H_{p})/s+1
\end{equation}

Where $s$ is the stride or step size and $H_{p~ out}\times W_{p ~out}\times D_{conv}$ is the size of the output of pooling layer, respectively. Generally, we may have more than 1 convolution and pooling layer, depending on the complexity of the problem and amount of available dataset.

The output of the feature extraction part, is fed to the
fully-connected layers, where it is passed through hidden layer's activation functions and finally a softmax activation function at the output of the last fully-connected layer, such that the class to which the input of the model belongs to is determined.
Note that the number of neurons at the output layer is equal to the number of classes. Equation \ref{softmax} shows how a class is detected at the output layer. 

\begin{equation}
   \arg \max _{0<k<c}  P_k(x) = f(X_i)
   \label{softmax}
\end{equation}

where, $c$ is the number of classes, $P_c(x)$ is the confidence score for predicting class $k$.  $X_i$ is the $i^{th}$ sample (input) and $f(.)$ is the softmax activation function. The class with the highest confidence score is the predicted class.

Having such various characteristics, CNN application has been extended to more challenging domains like time series prediction and classification.

\subsection{Time series classification (TSC)}

Combined with time series, deep learning frameworks can be used as a predictive tool, e.g. predicting neurological disorders like epileptic seizure[\citet{cogan2014wrist}], or neurological sicknesses like cybersickness
[\citet{islam2020automatic}], diagnostic method, like cardiovascular disorder detection [\citet{rajpurkar2017cardiologist}], or even therapeutic [\citet{liu2020research}] approaches in healthcare problems. 

Time series classification (TSC) is referred to the task of training a classifier on a set of time sequences, $X$, with the corresponding labels, $Y$, such that the trained classifier is able to predict the labels of a previously unseen test dataset [\citet{ebadi2019implicit}, \citet{he2015delving}].
 

Previous works have used different methods for TSC using distance measures, namely:
 euclidean distance (ED),  dynamic time warping (DTW) [\citet{ratanamahatana2005three}e] and edit distance with real penalty (ERP) [\citet{chen2004marriage}]. Further, \citet{keogh2003need} claimed that machine learning algorithms significantly outperform those traditional and empirical methods. 

\citet{kampouraki2008heartbeat} applied a support vector machine (SVM) method for heartbeat classification. \citet{chaovalitwongse2007time} used K-nearest neighbors (KNN) method to classify the brain abnormal activity. Moreover, random forest (RF), a machine learning algorithm that fits a number of decision tree classifiers on various sub-samples of the dataset, has been proposed to be a promising tool in TSC problems [\citet{deng2013time}]. Nonetheless, in order to be an accurate tool, all aforementioned machine learning methods need a feature extraction process before feeding the data into the model.

Due to having the ability of automatic feature extraction, CNN can solve this problem. Moreover, \citet{zeiler2014visualizing} has shown that 1D and 2D CNN could be a powerful in analyzing signals, where we have multi-modal time series.

\section{Materials and Methods}
\label{sec3 \protect}

\subsection{Dataset}

We used a publicly available data set for assessment of Neurological Status [\citet{goldberger2000physiobank}], comprising accelerometer in 3 dimensions (ACC), electrodermal activity (EDA) and body temperature (T) collected with a sample frequency of 8 [Hz]; heart rate (HR) and an estimate of the amount of oxygen in the blood calculated by oxygen saturation (SpO2) sampled at 1 [Hz]. Data was collected with wrist-worn sensors from 20 college students (6F/14M, age = 26.05 $\pm$3.8 yrs) undergoing various experiments with the purpose of probing a subject's response to different types of stress.

\begin{figure}[b]
 \centering 
 \includegraphics[width=\columnwidth]{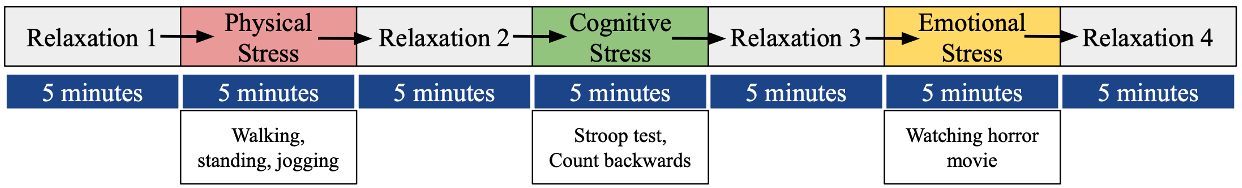}
\caption{Experiment design for data collection. Relaxation is used for helping the subjects get to the baseline after each stimulation phase. \protect }
 \label{data}
\end{figure}

The experimental protocol is shown in Fig \ref{data}

Five minutes were recorded for each of the following stages: relaxation, physical activity, cognitive stress and emotional stress. Relaxation is considered as the baseline, while physical activity is broken into three parts with different intensities: standing for one minute, walking for two minutes and finally jogging for two minutes. After another five minutes relaxation, the individuals experience a cognitive stress stimulation consisting in counting backward a list of
numbers for 3 minutes and then take the Stroop test for 2 minutes which is reading the names of colors written in different color ink and saying what color the ink is. Finally, after another five minutes relaxation, subjects are shown a a horror movie for five minutes to be emotionally stimulated.

In order to avoid redundancy, among all the relaxation phases, we only included  the first one into our computations. Moreover, the first 40 seconds of the cognitive phase, during which the subjects received explanations on the clinical protocol regarding the experiments, were considered as emotional phase, given that subjects showed patterns similar to the emotional patterns. 

We are going to apply a CNN classifier to discriminate the dataset into 4 phases, i.e Relaxation, Physical activity, Cognitive stress, Emotional stress.

\subsection{Data Preprocessing}

The data for each individual was separated into four phases: relaxation, physical activity, cognitive stress and emotional stress. Then the data for all subjects were concatenated along time axis.

To have the aggregated information of all three dimensions in one variable (time series), the magnitude of the 3D accelerometer signal was computed as:

\begin{equation}
\centering
    ACC= \sqrt{\sum_{i=0}^{l} {x_{i}^2 } + \sum_{i=0}^{l} { y_{i}^2 }+  \sum_{i=0}^{l}{z_{i}^2} }
\end{equation}

Where \textit{l} represents the length of signal.

It is a well known fact that EDA can be modeled as a sum of three terms: the phasic and tonic components plus an additive noise ( see Eq. \ref{cvx}) [\citet{greco2015cvxeda}].

The phasic component is an impulse-shaped signal describing the sudden changes in EDA signal, however,  tonic is a smooth-shaped sequence corresponding with the slow changes in EDA.

\begin{equation} 
\centering
    e_k= p_k + t_k +\epsilon_k
    \label{cvx}
\end{equation}

where $k$ is the time index,  $e_k$, $p_k$ and $t_k$ represent the eda, phasic and tonic component, respectively, and $\epsilon_k$ is a an independent and identically distributed (iid) signal with mean zero and variance 1 (Gaussian Random Variable), representing the additive noise term.  Note that all terms, $e_k$, $p_k$ , $t_k$ and  $\epsilon_k$ are the same length column vectors.

In addition to these three variables, HR and SpO2 variables are also taken into account for our computations. Since we are dealing with a multi-variate dataset, to solve the problem of having various distribution for each variable, we use Z-score method [\citet{kellaway1969advanced}] for each of them separately:

\begin{equation} 
\centering
\label{zscore \protect}
    Z= \frac{X_i - \mu_i }{\sigma_i}
\end{equation}

where $X_i$ is the variable, $\mu_i$ and $\sigma_i$ are the associated mean and the standard deviation of the variable, respectively, and $i={1,...,,25}$ is the number of variables, which in our case is 25.

Moreover, we took the first, second and third derivative of each variable in time domain separately, and stacked the resulted variables below the time-domain components. 

Furthermore, prompted by a previous study of our group  [\citet{eda_abs}], showing that there are relations between changes in neurological status and frequency components of simultaneously recorded physiological signals, we computed the frequency component, the absolute magnitude of the short time Fourier transform (STFT) (Eq. \ref{stsft}), for each time series to capture the signal’s frequency content over time. Then we stacked the resulted 5 variables in frequency domain, below the time domain variables.

\begin{equation} 
\centering
\label{stsft}
    X(\tau, \omega)  = \int_{-\infty}^{\infty} x(t) w(t-\tau) e^{-i\omega t } dt
\end{equation}

where $w(\tau)$ is the window function, $x(t)$ is the signal in time domain and $X(\tau, \omega)$ is the Fourier transform representing the magnitude and phase of $x(t)$ over time and frequency. $\tau$ and $\omega$ are time and frequency index, respectively. It should be mentioned that we only take magnitude of $X(\tau, \omega)$ as the frequency component of each variable.  
As a result, we acquired a 25 variable dataset, consisting of initial time domain components, frequency components and first, second and third derivatives of time domain variables.

Finally, to make the data ready for being fed into the CNN model, we used sliding windows of width 25 and length 30, equal to the number of variables, with 50\% overlap and moved it all along the dataset to create samples of size 25 $\times$ 30.

To compare the effect of each sets of variables, we created datasets with different number of variables such that: dataset one only includes the raw signals in time domain, dataset two includes time and frequency components. The third, fourth and fifth datasets includes components of time domain along with the first, second and third derivatives, respectively. Dataset six includes time, frequency and first derivatives.The seventh one consists of the time, frequency, and second derivatives and finally the eighth one comprises all 25 variables.

\section{Results}
\label{results \protect}
\subsection{1D CNN}
After training the model with all parameters in Table \ref{table:activation}  
the best results were acquired by using ELU with $a=1$ as the activation function of the convolution layers, since it diminishes the vanishing gradient effect, and TanH for hidden layers of the neural network. Moreover, Root Mean Square Propagation (RMSprop) with learning rate of 0.001 and the Categorical Cross Entropy was selected as the optimization algorithm and the loss function, respectively.

Since we had 25 variables, to know the effect of each set of them on our classification results, i.e, dataset one to eight, we trained the 1D-CNN for all datasets separately.

To make sure that the training, validation and test datasets are representative of the overall distribution of the dataset, we apply shuffling such that all training, validation and test datasets may include information from all subjects. 

Then, We split the data into 70\% for training, 10\% for validation during training and 20\% for the test dataset.
Table \ref{1d-various-variables} shows the training results of the 1D CNN for 100 epochs and batch size of 32, for different datasets. 
Accuracy, area under the curve of the receiver operation characteristic (AUC-ROC) and area under the recall-precision curve (AUC-RP) are used as the evaluation metrics to compare the results. We can see that using all 25 variables can significantly improve the results, hence we will use the dataset number eight.

\begin{table}[h]
\begin{center}
    
\caption{Results of training 1D CNN  with different combinations of components \protect}
    \label{eval-all-data\protect}
    \label{1d-various-variables\protect}
    \begin{tabular}{cccc}
    Dataset &  Accuracy (\%)  & AUC-ROC & AUC-RP \\\hline
    One & 95.14 & 0.968 &0.979  \\
    Two & 97.0 & 0.978 & 0.984 \\ 
    Three & 96.45 & 0.975 & 0.983 \\ 
    Four  & 95.31 & 0.969 & 0.980 \\ 
    Five  & 95.2 & 0.998 & 0.979\\
    Six  & 97.36 & 0.978 & 0.983 \\
    Seven  & 98.05 & 0.985 & 0.987 \\
    Eight & \textbf{99.82} & \textbf{0.999} & \textbf{0.998} \\
\hline
\end{tabular}
\end{center}
\setlength{\tabcolsep}{1pt}
\end{table}

\begin{figure}[b]
\centering
     \includegraphics[height=2in]{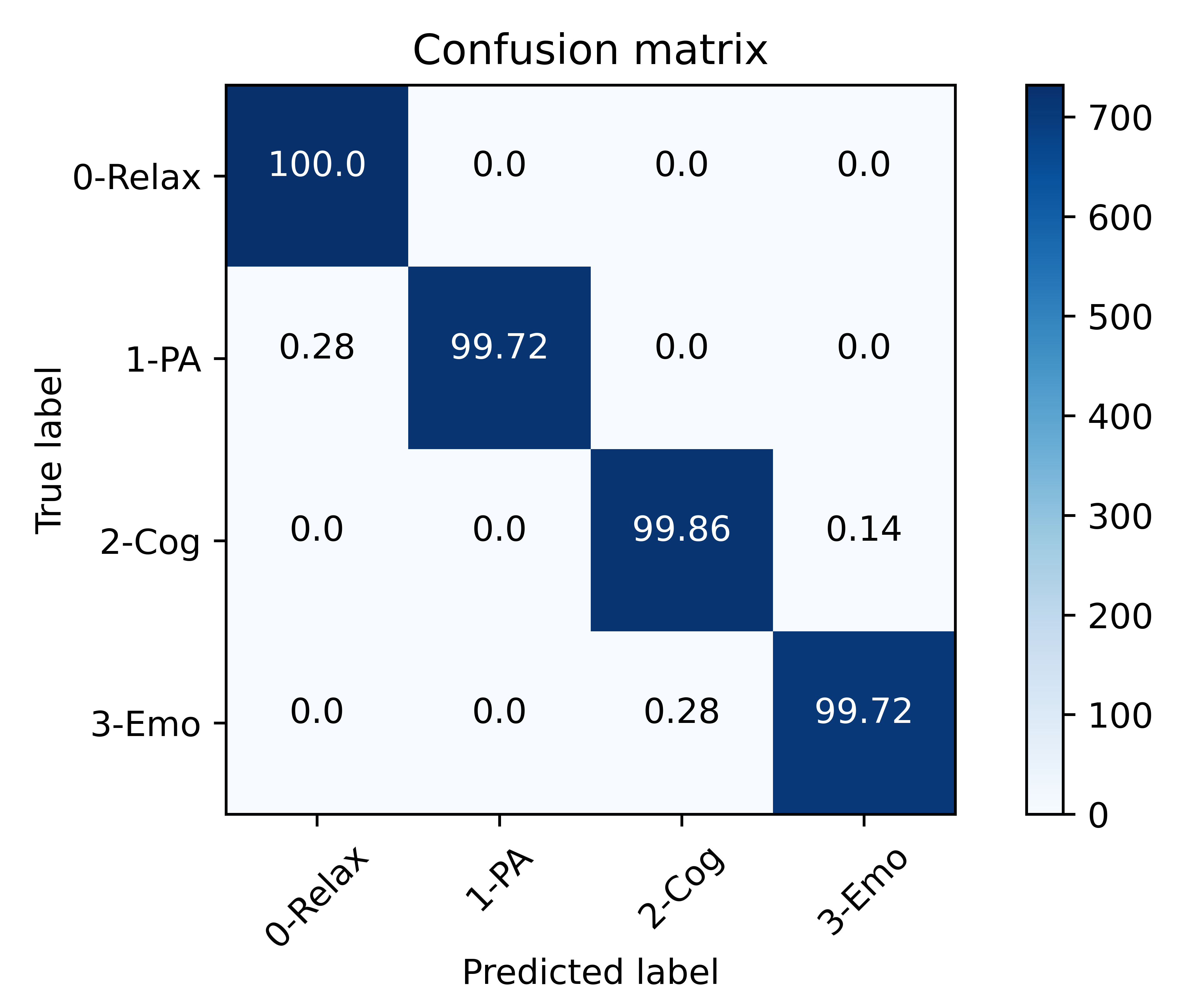}

 \caption{Confusion matrix for 1D CNN model (\%)\protect}
 \label{1dcnn-conf-all}
\end{figure}

Table \ref{1d-cnn-result} demonstrates the performance of the 1D CNN model evaluated with other metrics, namely: precision, recall and F1-score of each classes as well as the total classification accuracy.
Moreover, figure \ref{1dcnn-conf-all} shows the confusion matrix.

\begin{table}[h]
\begin{center}
\caption{Evaluation results of the 1D CNN model \protect}
\label{1d-cnn-result}
\begin{tabular}{cccc}
Class& Precision (\%)  & Recall (\%)  & F1-Score  (\%) \\\hline
Relaxation & 99.72 & 100.00 & 99.85 \\
Physical Activity  & 99.72 & 99.72 &99.72  \\ 
Cognitive Stress  & 99.72 & 99.86 & 99.78 \\ 
Emotional Stress  & 99.85 & 99.72 & 99.78 \\ 
\hline
&Accuracy  &\textbf{99.82}\% \\
\hline
\end{tabular}
\end{center}
\end{table}

\subsection{2-Dimensional CNN}
 
\begin{figure}[b]
\centering
     \includegraphics[height=2in]{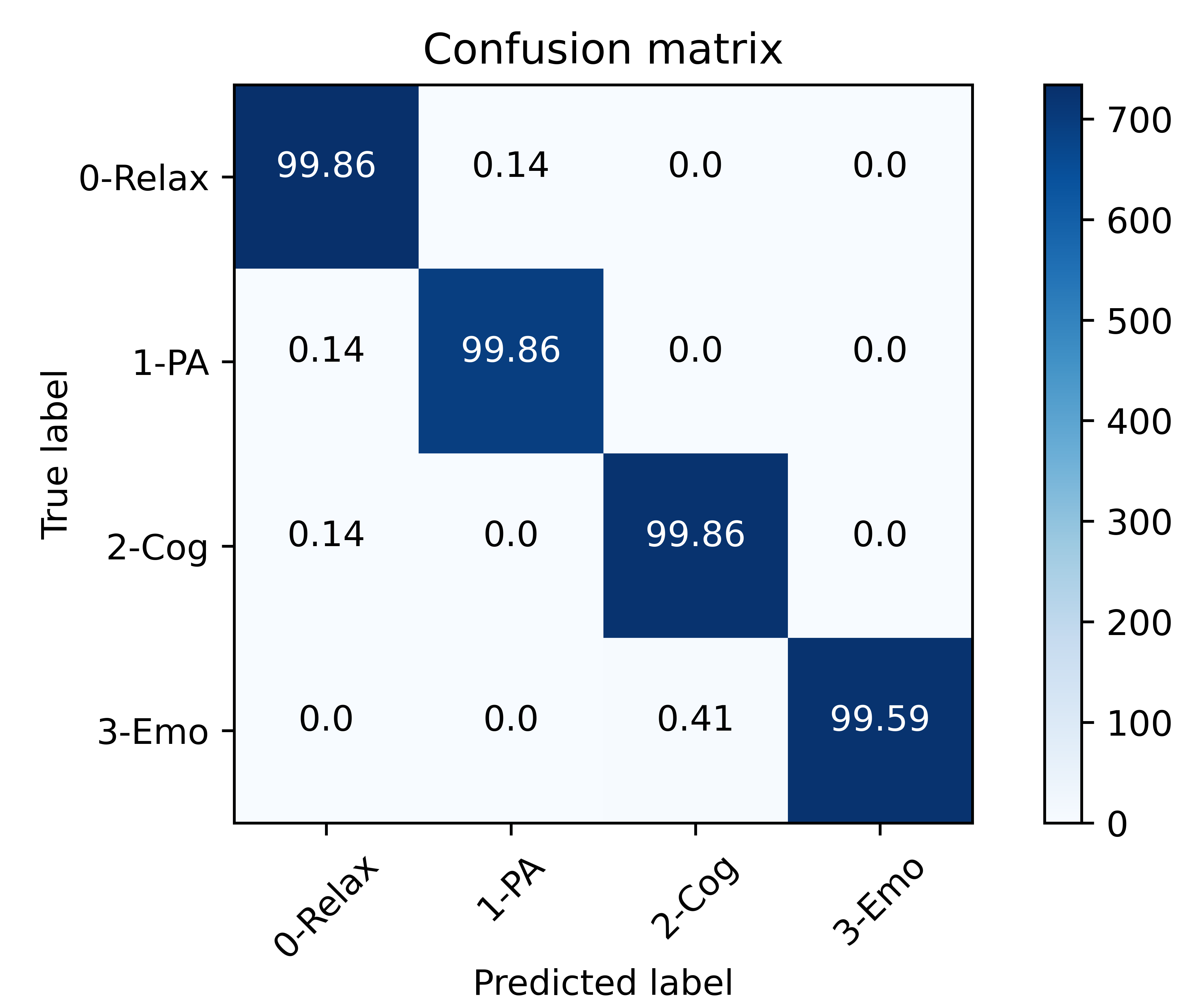}
     
 \caption{Confusion matrix for 2D CNN model (\%)) \protect}
 \label{fig-2dcnn}
\end{figure}

We also implemented a 2D CNN using the same 
hyper parameters and split size as we used for 1D CNN.
Results of training after 150 epochs, using early stopping point and with batch size of 32 is shown in Table \ref{recal-2d}.
Moreover, figure \ref{fig-2dcnn} shows the confusion matrix of 2D CNN for classifying the whole dataset of size 25 variables into the 4 neural stages.

\begin{table}[t]
\begin{center}
\caption{Evaluation results of the 2D CNN model \protect}
\label{recal-2d}
\begin{tabular}{cccc}
Class& Precision (\%)  & Recall (\%)  & F1-Score  (\%) \\\hline
Relaxation & 99.72 & 99.86 & 99.78 \\
Physical Activity  & 99.86 & 99.86 & 99.86 \\ 
Cognitive Stress  & 99.59 & 99.86 & 99.72 \\ 
Emotional Stress  & 100.00 & 99.59 & 99.79 \\ 
\hline
&Accuracy  &\textbf{99.61}\% \\
\hline

\end{tabular}
\end{center}
\end{table}

\subsection{Comparison with SVM and RF}
To better evaluate the benefits of CNNs in the neurological status classification task at hand, we trained a multi-class SVM and a RF model and subsequently compared the results obtained to those of CNNs.

Since for training these models we need to apply a feature extraction method prior to training the model,
the following statistical features were extracted from each of the windows (samples) of the dataset: \textit{mean, maximum, minimum, range} (the amount of changes for each variable from beginning to the end of each window) and the \textit{first derivative}.

For training a multi-class SVM, after creating the parameter grid based on the results of random search, the best parameters were obtained using Grid Search method such that, the regularization parameter, C, was set to 100, gamma value 0.001 and the Radial Basis Kernel (RBF) was selected.

Moreover, the same features were used for training a RF model with complexity parameter of zero, 150 random state, number of estimator of 130 and the gini criterion. 

As Table \ref{comparison} shows the classification accuracy and F-1 score resulted from each method, we can see that 1D and 2D CNN have almost the same results, while due to having less trainable parameters, 1D CNN is significantly faster than 2D CNN., hence it can be a better choice in this case.

\begin{table}[h]
\begin{center}
\caption{Comparing CNN with SVM and RF \protect}
\label{comparison}
\begin{tabular}{ccc}
Classifier& Accuracy (\%)  & F1-Score  (\%) \\\hline
SVM & 84.45 & 84.25 \\
RF  & 92.29 & 92.50 \\ 
1D CNN  & \textbf{99.82}  &99.78 \\ 
2D CNN  & 99.61 &  99.77 \\ 
\hline

\end{tabular}
\end{center}
\end{table}

\subsection{Adding Noise}

In order to challenge our proposed approach, we added a white noise with zero mean and standard deviation 1 to all the sequences and trained all previously mentioned models again. Table \ref{nose-results} shows the performance of all methods resulted after training with the noisy dataset. It is observable that CNN models significantly outperforms SVM and RF in presence of noise. However, it should be noticed that in this case, 2D CNN outperforms 1D CNN which is because of having 2D kernels, i.e., 2D CNN has more trainable parameters and can learn more detailed features from the dataset, hence it can be more powerful than 1D CNN in complex problems. 
 
\begin{table}[h]
\begin{center}
\caption{Comparison results after adding noise to the dataset \protect}
\label{nose-results}
\begin{tabular}{ccc}
Classifier & Accuracy & F1-Score (\%) \\
\hline
SVM  & 70.25 & 65.75 \\
RF  & 71.84 & 69.23\\ 
1D CNN  &  93.54 & 93.47 \\ 
2D CNN  &  \textbf{97.46} & \textbf{97.32} \\ 

\hline

\end{tabular}
\end{center}
\end{table}

\section{Summary and Conclusion}
\label{conclusion \protect}

In this paper we exploited deep learning approach to solve classification problems in a multi-modal physiological time series dataset recorded from wrist-worn sensors. After some preprocessing steps, we obtained 99.82\% and 99.61\% classification accuracy as well as 99.78\% and 99.77\% F1-score for 1D and 2D CNN models, respectively. As for the comparison, we trained two well-known machine learning algorithms in classification problems, multi-class SVM and RF, using our manually extracted features. The results prove that both 1D and 2D CNN models outperform SVM and RF method, however, due to being computationally more affordable, 1D CNN algorithm is faster than 2D CNN. Moreover, to make it more challenging, we added a white noise with zero mean and variance 1 to the dataset and trained all the models again. This time also, both CNN models outperformed SVM and RF, even more significantly, while due to having 2D convolution kernels and consequently more trainable parameters, 2D CNN performed better than 1D CNN.
Hence we showed that the robustness to noise characteristics of the CNN models in comparison to the  machine learning algorithms, makes them very useful tools in working with noisy datasets, like physiological time series or medical datasets. Moreover, we could see that CNN models generate features general enough to show promising performance on previously unseen test dataset.

Our future work will be devoted to leveraging these methodology on more challenging physiological datasets, and also using this methodology for other purposes like designing a predictive or prohibiting device, informing the patients prior to having high levels of stress which may put them into unpleasant health conditions.

\bibliography{reference}    
\end{document}